# MODELING THE DYNAMICS OF SOCIAL NETWORKS

Victor V. Kryssanov, Frank J. Rinaldo
*Faculty of Information Science and Engineering, Ritsumeikan University, Kusatsu, Shiga, Japan*
kvvictor@is.ritsumei.ac.jp, rinaldo@is.ritsumei.ac.jp

Evgeny L. Kuleshov
*Department of Computer Systems, the Far-Eastern National University, Vladivostok, Russia*
kuleshov@lemoi.phys.dvgu.ru

Hitoshi Ogawa
*Department of Information and Communication Science, Ritsumeikan University, Kusatsu, Shiga, Japan*
ogawa@airlab.ics.ritsumei.ac.jp

Abstract:   Modeling human dynamics responsible for the formation and evolution of the so-called social networks – structures comprised of individuals or organizations and indicating connectivities existing in a community – is a topic recently attracting a significant research interest. It has been claimed that these dynamics are scale-free in many practically important cases, such as impersonal and personal communication, auctioning in a market, accessing sites on the WWW, etc., and that human response times thus conform to the power law. While a certain amount of progress has recently been achieved in predicting the general response rate of a human population, existing formal theories of human behavior can hardly be found satisfactory to accommodate and comprehensively explain the scaling observed in social networks. In the presented study, a novel system-theoretic modeling approach is proposed and successfully applied to determine important characteristics of a communication network and to analyze consumer behavior on the WWW.

## 1 INTRODUCTION

There is an increasing number of reports that human behavior underlying the development of social (communication, entertainment, financial, and the like) networks does not follow the Poisson statistics conventionally employed to describe an individual's bursty activities (Sheskin, 1997) but, instead, reveals the scaling dynamics conforming to the power law (Johansen, 2004; Barabasi, 2005; Oliveira and Barabasi, 2005; Adamic and Huberman, 2000; Scalas *et al.*, 2006). Striving to understand the mechanisms and factors responsible for the scale-free behavior, researchers have been quick to affiliate the dynamics of social networks with the familiar Zipfian phenomena (see Newman, 2005, for a general survey, also Barabasi and Albert, 1999).

There exist a rich variety of stochastic processes leading to a power, heavy-tailed (e.g. Zipf, Zipf-Mandelbrot, or Pareto) form of the probability distribution of an observed random variable (Mitzenmacher, 2003). Only a small fraction of these processes, however, would be considered relevant to discuss in a social or anthropological context peculiar to the development of social networks. Even fewer processes have actually been explored as possible generating mechanisms for the network dynamics and tested against real-world data.

Adamic and Huberman (2000) gave an explanation for the power-law distribution of the consumer activities in a global e-market, such as the World-Wide Web (WWW). The proposed model exercises the well-studied multiplicative growth stochastic mechanism for the network expansion but carries no implication about the human behavior. Barabasi (2005) suggested a version of the preferential selection mechanism to describe the dynamics observed in a university e-mail network. While he did propose a model for human communicative behavior, which is, effectively, choice based on priorities, this model requires making rather implausible assumptions (e.g. about uniformly distributed priorities) and yet demonstrates poor predictive results even for the data originally used in the study (see Figure 1; also Stouffer *et al.*, 2005). Johansen (2004) derived an empirical formula, which provides a good approximation for the general response rate of a human population, working with the same data as the previous author (Figure 1). Another example of the empirically grounded approaches to modeling the dynamics of social networks is a modification of the Zipf-Mandelbrot law – the formula suggested by Krashakov *et al.* (2006) to characterize the popularity of Web-sites that apparently has a predictive power better than the classic (e.g. the "pure" power or Zipf law) models. The latter two studies, however natural, offer little insight on why the observed networks exhibit scale-free properties.

In the absence of a sufficiently universal alternative to the power law (see Solow *et al.*, 2003, for a relevant discussion), the above mini-survey is quite indicative of the current situation with the understanding and modeling of the dynamics of social networks. Whenever the true mechanism underlying the observations is not known, the most probable scenario is that any process generating heavy-tailed data is either "by default" (i.e. with a minimal, if any, attention to statistical hypothesis testing and model validation) attributed to (a version of) one of the well-studied power-law generating mechanisms, such as multiplicative growth, preferential attachment, optimal coding, etc. or simply approximated with an empirical "*a la* Zipf" formula having an arbitrary interpretation that can hardly be discussed in a context different from mere curve fitting.

In the presented study, the authors aim to improve upon this, in essence theoretical, deficiency and focus on the development of a reasonably universal approach that would provide a distinct modeling perspective and have a potential to deliver a plausible and verifiable explanation of scale-free phenomena discovered in diverse social networks.

The next section gives a general mathematical framework. It is applied to analyze possible reasons of the power law patterns in the observed behavior of complex systems. Two experiments are then conducted to determine the dynamic structure of social networks, based on the proposed theory, and their results are briefly discussed. The study's conclusions are drawn, and plans for future research are outlined.

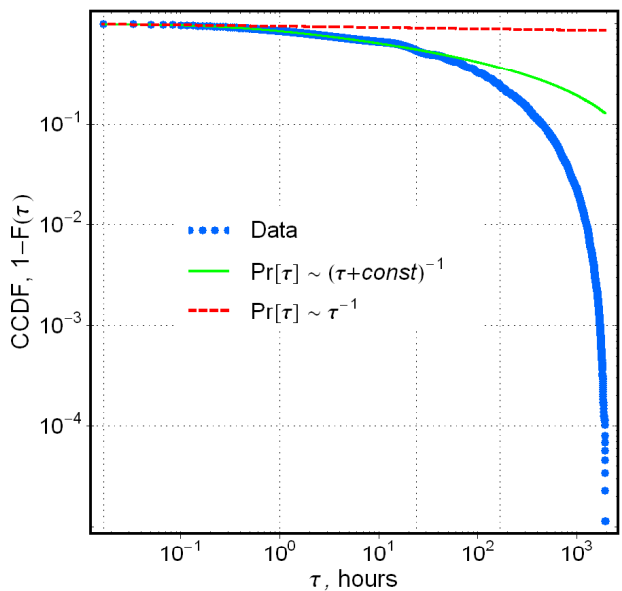

Figure 1: Problems with explaining the heavy-tailed activity pattern in e-mail communications: While the empirical formula (solid line) derived by Johansen (2004) provides a better approximation for the distribution of $\tau$, the time taken by an individual to reply to a received e-mail, than Barabasi's model (dashed line) based on an activity-prioritizing mechanism (Barabasi, 2005), it gives no clue about the generating process. (For details on the data, see Eckmann *et al.*, 2004.)

## 2 SYSTEM-THEORETIC FRAMEWORK

In this section, we will analyze the observed behavior of the so-called complex systems – the conglomerates (physical, social, cognitive, cybernetic, or the like) whose internal mechanisms and structure by some reason cannot be inspected in full. Power-law phenomena are very often associated with (produced by, observed in, etc.) such systems.

Let us consider a system $\Omega$ defined in a very general sense, i.e. as the object of investigation (not necessarily physically grounded). An observable $O$ is a property of the system $\Omega$ that can be investigated in a given context. We will assume that $\Omega$ exists in different states and that different states of the system release themselves as different outcomes of observations (measurements, etc.) associated with observables $O$. The latter means that the system states (or behavior, seen as state change) are in principle conceivable through their representations resulting from observations of $O$.

We will seek to determine the distribution of the occurrence number of different representations associated with a given observable across increasing expenditures of time. In so doing, we will assume that *a*) the same state can have different representations, and *b*) different states can have the same representation. The analysis will be in three steps.

**Step I**: to characterize the occurrence number (or rate) of different representations of one (identical) state for the same observable $O_0$.

<u>Case 1</u>: Let the process of system state representation implement an efficient encoding procedure so that $\bar{k}_0 t_r = const$, where $\bar{k}_0$ is the expectation of a discrete random variable $K_0$ revealing the occurrence number of different representations, and $t_r$ is the average time of state representation.

To estimate $f_{K_0}(s)$ the probability mass function (PMF) of $K_0$, we will maximize its entropy $H = -\Sigma_s f_{K_0}(s) \ln f_{K_0}(s)$, $s = 1, 2, \ldots$, subject to constraints $\Sigma_s f_{K_0}(s) = 1$ and $\Sigma_s s f_{K_0}(s) = \bar{k}_0$. This will give us "the least biased estimate possible on the given information" (Jaynes, 1957).

The optimization problem is solved using a Lagrangian approach. The Lagrangian function is defined as

$$\mathcal{L}(f_{K_0}) = -\sum_s f_{K_0}(s) \ln f_{K_0}(s) + \gamma(1 - \sum_s f_{K_0}(s)) + \lambda(\bar{k}_0 - \sum_s s f_{K_0}(s)), \quad (1)$$

where $\gamma$ and $\lambda$ are coefficients, with optimality conditions $\partial \mathcal{L} / \partial f_{K_0}(s) = 0$, $\partial \mathcal{L} / \partial \gamma = 0$, and $\partial \mathcal{L} / \partial \lambda = 0$ (Cover and Thomas, 1991). From the first and second conditions, one can then derive

$$f_{K_0}(s) = \Pr[K_0 = s \mid \lambda] = (e^\lambda - 1) e^{-s\lambda}, \quad (2)$$

where $s = 1, 2, \ldots$. Since $\bar{k}_0 = \sum_{s=1}^\infty s f_{K_0}(s) = e^\lambda / (e^\lambda - 1) = 1/f_{K_0}(1)$ and $\bar{k}_0 \gg 1$, $f_{K_0}(1) \ll 1$ and $\lambda \ll 1$, and therefore $\lambda \approx 1/\bar{k}_0$.

<u>Case 2</u>: Let us now consider a different-type system and impose a conservation constraint for the representation (observation) time by requiring that at any time, only one but not necessarily the same property of the system is evaluated. In other words, we will assume that there are multiple competing observables for the same state. We will also assume that these observables are independent. (To simplify technicalities, the following discussion will mainly be built around the continuous case, i.e. for $k_0$ the continuous counterpart of $K_0$, yet with the customary abuse of the notation when the same symbol is used to refer to a random variable and to its value.)

For $\theta$ a period of time, $w_0$ the rate of the representation change is given as $w_0 = k_0 / \theta$. Under the above assumptions, the dynamics of $w_0$ can be modeled using a system of differential equations defined as follows:

$$\frac{dw_i}{dt} = a_i \left( \rho\mu - \sum_{n=0}^N w_n \right) + \eta_i \quad (3)$$

where $\mu$ is the investigated state rate (characterizes the "true," as opposed to the observed, behavior of the system), $N$ is the number of competing observables, $i = 0, 1, \ldots, N$; $\eta_i(t)$ is a Gaussian noise (e.g. due to measurement errors) with average zero, some $a_i(t) > 0$, and $\rho$ is a parameter to account for the efficiency of the representation process (i.e. the system state may principally be only to an extent available for observation).

Equations (3) describe a diffusion process in the vicinity of a hyperplane $\rho\mu = \Sigma_n w_n$ formed by $N+1$ observables with $w_n$ representation rates, whose values are (approximately) uniformly distributed in the interval $[0, \rho\mu]$. Due to the hyperplane condition, there can be only $N$ mutually independent observables, say $O_1, \ldots, O_N$. $\Sigma_n k_n = \Sigma_n w_n \theta = \theta\rho\mu$ by definition. The probability that $k_n \geq k_0$, $n = 1, \ldots, N$, can be calculated as a product of the marginal distributions $f(k_n) = 1/\theta\rho\mu$, that then yields $\Pr[k_1 \geq k_0, \ldots, k_N \geq k_0] = (1 - k_0/\theta\rho\mu)^N$. Probability theory defines the cumulative distribution function (CDF) for some $x_0$ taken from the set of all random variables that obey a given probabilistic law as $F(x_0) = \Pr[x_1 < x_0, \ldots, x_N < x_0]$. In this context and for $N \gg 1$, one can write:

$$F(k_0) \propto 1 - (1 - k_0/\theta\rho\mu)^N \approx 1 - e^{-k_0 \lambda}, \quad (4)$$

where parameter $\lambda = t'_r(\theta\rho)^{-1}$ is, up to the constant $1/\theta\rho$, determined by $t'_r = (\mu/(N+1))^{-1}$ the representation time averaged over the observables (note that generally, $t_r$ of Case 1 is not equal to $t'_r$). At this point, we would like to note that while there would be a number of modeling scenarios both, similar to and different from those of Case 1 and 2, which would produce an exponential form for the distribution sought, the approaches discussed above have two important implications. First, the models defined with Equations (2) and (4) both stipulate that under other similar conditions, more often met representations correspond, on average, to system states with a shorter representation time. Second, for an occurrence number significantly greater than 1, the parameter $\lambda \ll 1$.

**Step 2**: to characterize $k(t)$ the representation occurrence number of many different states for the same observable $O_0$.

For the system $\Omega$, the measured stochastic variable $k$ will be a sum of random variables $k0_0 + k1_0 + k2_0 + \ldots$, where the summands are due to different states having identical representations. The statistical properties of $k$ will then depend on the parameter $\lambda$ that can naturally vary (e.g. as a result of a variation in the average representation time for different states). Let $g(\lambda)$ be the probability density function (PDF) of $\lambda$. For a large number of states investigated by means of $O_0$, $f(k)$ the PDF of $k$ is defined as a $g(\lambda)$ parameter-mix of $f(k | \Lambda = \lambda)$:

$$f(k) = f(k | \Lambda = \lambda) \bigwedge_\Lambda g(\lambda) = \int_0^\infty g(\lambda) P(k_0) d\lambda, \quad (5)$$

where $P(k_0)$ is the PDF of $k_0$ discussed in Step 1.

**Step 3**: to generalize the result of Step 2.

The random variable $k$ may reflect more than one (presumably associated with the observable $O_0$) property of the system $\Omega$, while the system mechanisms controlling the observable may be heterogeneous in time (e.g. owing to environmental perturbations). This can provoke the existence of more than one probability distributions for $\lambda$. When $M$ the number of statistically independent factors influencing the observation (or the system behavior) is finite, $P(k)$ the PDF of the occurrence number of system state representations can be estimated as

$$P(k) = \sum_{i=1}^M c_i f_i(k), \quad (6)$$

where weights $c_i$ give the likelihood to observe the influence of the $i$-th factor on the random variable $k$, and each $f_i(k)$ is specified with Equation (5) and is determined by the (sub)system parameters as it was discussed in Step 1.

## 3 WHEN THE POWER LAW?

The analytic framework formulated in the previous section is fairly general and can be applied to analyze the behavior of virtually any complex system. It should be emphasized however, that the focus of the developed model is on the frequency (count) of observed activities rather than on their durations. Most of the modeling approaches discussed in the introduction are therefore not directly comparable to the one proposed in this study. Given the fact that in social systems, there often exists a detectable (though not always easily formalizable) connection between the frequency of a certain activity and its duration, it appears interesting to explore under what conditions Equation (6) may produce a power form of the probability distribution.

An acute reader would have already noticed that the substitution of the exponential PDF into Equation (5) yields a Laplace transform of the product $\lambda g(\lambda)$:

$$f(k) = \int_0^\infty \lambda e^{-k\lambda} dG(\lambda), \quad (7)$$

where $G(\lambda)$ is the CDF of $\lambda$.

This is a very nice result since it, owing to Bernstein's theorem (Bernstein, 1928), stipulates that if $f(k)$, the PDF of the observed variable, is completely monotone, i.e. all its derivatives exist and $(-1)^n f^{(n)}(k) \geq 0$ for any integer $k > 0$ and $n \geq 1$, there can always be found some proper $G(\lambda)$ in effect describing the internal (i.e. not directly observed, "true") dynamics of the system. There is a large class of probability distributions for $\lambda$ (e.g. originated from or simply approximated with the Beta of the Second Kind probability distribution $g(\lambda) = (\Gamma(p+q)/\Gamma(p)\Gamma(q))\lambda^{p-1}(1+\lambda)^{-(p+q)}$, where $\Gamma(\cdot)$ denotes the Gamma function, $p > 0$ and $q > 0$ are parameters, which encompasses many commonly used distributions, such as the Log-Normal, Gamma, Weibull, etc. – see McDonald and Xu, 1995) that will cause "fat" tails of the observed data $f(k) \propto k^{-\beta}$ for some $\beta > 0$ as $k \to \infty$ (Abate and Whitt, 1999). This asymptotic property will, however, not necessarily be maintained for small $k$.

Generally, the developed model dictates that in the case of a homogeneous (i.e. assuming the existence of one PDF $g(\lambda)$) system, candidate distribution functions for the description of the statistical properties of $k$ and $\lambda$ should satisfy Equation (7). In view of this, an interesting exercise would be to find a family of probability functions satisfying the Laplace transform (7). Unfortunately, no closed analytic forms for $g(\lambda)$ exist in many cases of long-tailed $f(k)$.

Equation (6) further stipulates that in order for $P(k) \propto k^{-\beta}$, either all the summands should have an identical "heavy-tailed" form (that would indicate certain self-similarity existing in the system) or at least one of the summands should have a power form with an exponent $\beta_i$ small enough to dominate the asymptotic behavior of the other distributions.

## 4 MODELING THE DYNAMICS OF SOCIAL NETWORKS

To verify the proposed model against real-world data and explore its predictive and analytic capabilities, two experiments have been conducted.

### 4.1 Experiment 1

Data used in the experiment is a sample representing the timing of e-mails sent and received by a group of ~10000 people at a university in Europe during a period of ~80 days. The corresponding server log-file was obtained from the authors of Reference (Eckmann *et al*., 2004). Specifically, we have focused on the time taken by an individual to reply to a received message – the human response rate; there have been extracted ~24000 of reply times from the file. It was reported elsewhere (Eckmann *et al*., 2004; Johansen, 2004; Barabasi, 2005) that power-law generating mechanisms could be behind the formation of the distribution of this data, as it exposes the characteristic (yet noisy) heavy-tailed pattern (see Figure 1; also the inset in Figure 2).

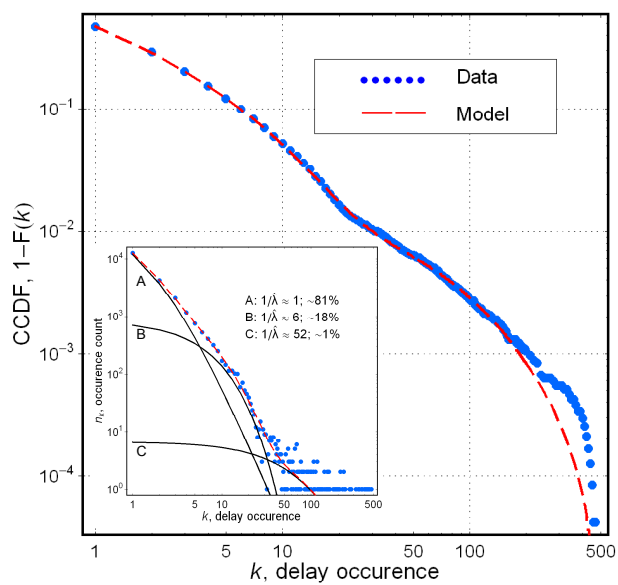

Figure 2: Estimation of the dynamic social structure based on an analysis of the traffic in a university e-mail network.

The discretization time interval for the delays with reply was set to 1 minute. No other preprocessing has been done. The investigated system in this case is the social system that existed at the university, and the observed property is the perturbed (by incoming e-mails) communication timing. It is expected that delays with reply to received e-mails reveal the rate of the system internal state change (e.g. in its simplest form, from "e-mail is not replied" to "e-mail replied").

The two-parameter Gamma distribution $g(\lambda) = b^\nu \lambda^{\nu-1} e^{-b\lambda} / \Gamma(\nu)$, where $\lambda \geq 0$, $b > 0$, $\nu > 0$, and $\Gamma(\cdot)$ denotes the Gamma function, was chosen to characterize the system's hidden dynamics, because this is a simple form well describing cognitive processes and "mental" reaction time (Luce, 1986). This form can, and possibly should, be considered for $g(\lambda)$ regardless what is the "true" mental architecture triggering one or another investigated human activity (van Zandt and Ratcliff, 1995). Taking into account the discrete nature of the observed variable value count and after substitution of the corresponding PMF and PDF into Equation (5), Equation (6) is specialized to

$$P(k) = \sum_{i=1}^{M} c_i \left( \frac{b_i^{\nu_i}}{(k-1+b_i)^{\nu_i}} - \frac{b_i^{\nu_i}}{(k+b_i)^{\nu_i}} \right), k=1,2,\ldots \ (8)$$

that is thus the probability mass function of the occurrence of a human response rate. Note that each of the summands has the form of the discrete Lomax distribution.

Figure 2 depicts the complementary cumulative sums calculated for the data and the model (8) with parameters $c_i$, $b_i$ and $v_i$ obtained by a numerical maximum likelihood method and $M = 3$ as yielding the smallest value of Akaike's Information Criterion, $\text{AIC} = -2\log(L(\hat{\phi}|x)) + 2n$, where $\log(L(\hat{\phi}|x))$ is the log-likelihood maximized with parameters $\hat{\phi}$ (for the PMF (8), $\hat{\phi} = \{\hat{c}_{i\neq 1}, \hat{b}_i, \hat{v}_i\}$, $i = 1,...,M$) for a given sample $x$, and $n$ is the number of estimable parameters (for $M = 3$, $n = 8$). AIC is a fundamental measure assessing the relative Kullback-Leibler distance between the fitted model and the unknown true mechanism, which actually generated the observed data (Akaike, 1983). Taking into account the information known from the original report (Eckmann *et al.*, 2004) about the structure of the social system in focus, models with $M$ the number of components ranging from 1 to 4 have been tried in the experiment. The second-best model had $M = 2$, $n = 5$ and was therefore simpler, but with an AIC value by 90 greater than in the case of $M = 3$ it had to be omitted from further consideration (Sakamoto *et al.*, 1986).

## 4.2 Experiment 2

To explore the dynamic structure of an e-market on the World-Wide Web, a data sample representing the activity of America Online (AOL) users (acting as consumers of the services provided by Web sites) has been obtained from the authors of Reference (Adamic and Huberman, 2000). The sample covers approximately 120000 sites accessed by 60000 users during one day.

Figure 3 shows the results of the modeling of the consumer activity dynamics with formula (8). It is assumed that a hit to a particular site corresponds to a specific mental or "goal" state, and that these states are common (i.e. shared) within the population (from a generic anthropological viewpoint, this seems a natural assumption). As no *a priori* information was available on the structure of the social system examined, four prediction models have been probed by varying the value of $M$ from 1 to 4. The two models displayed in the figure are statistically justified by the data and perform practically equally well.

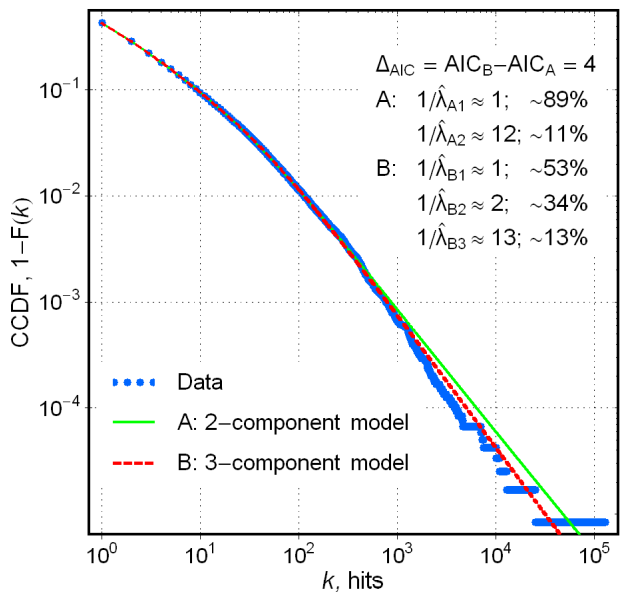

Figure 3: Modeling the Web site (server) visiting rate (hits) observed on December 1, 1997, in a segment of the WWW. (For details on the data, see Adamic and Huberman, 2000.)

## 5 DISCUSSION

It is quite illustrative that while a "pure" form of the power law would fail to reasonably accurately predict probabilities for the entire ranges of the data used in the experiments, as the corresponding complementary cumulative sums visibly do not form single straight lines on the double-logarithmic plots, the model specified with Equation (8) produces sound fits.

For the university e-mail network, Pearson's $\chi^2$ test does not reject the model with a significance level $\alpha = 0.001$ that might be considered good enough in the case of noisy data. An objection would, however, be made that the proposed model overfits the data: the large number of its parameters creates a situation when the fit may be driven by the random fluctuations rather than by the "true" statistical properties of the data.

The values of $c_i$ the parameters obtained in the first experiment suggest that the examined social system has an internal structure: there are two subsystems with different dynamics responsible for the generation of approximately 81 and 18% of the observed variety in delays with reply to an e-mail; about 1% of the occurrences – for the most typical and shortest delays – are probably caused by factors other than social (e.g. owing to an auto-reply function of the e-mail clients or the processing of long mailing lists) and may be excluded from consideration.

The larger, "static" subsystem – A in the inset of Figure 2, where the distributions are built for the data, the model, and the model's 3 components – produces on average longer yet unique delays (for a gamma-distributed $\lambda$, the estimate of its mean $\hat{\lambda} = \nu/b$). The second subsystem – B – is approximately 4 times smaller (or 4 times observationally less influential) and 6 times more dynamic (and hence, as it could be speculated, is more constrained and/or has stronger social ties). These size estimates principally conform to the ones reported in the original work (Eckmann *et al.*, 2004) and independently obtained through somewhat intricate analysis of the individual communications present in the sample. This, along with the fact that the obtained parameters behave just as it is implied by the model (back to Section 2), can be considered as strong evidence in support of the hypothesis that Equation (8) does describe the system behavior but not merely approximates the data.

In the second experiment, the models with two and three Lomax-distributed components are not rejected by Pearson's test with $\alpha = 0.1$. The difference in AIC calculated for these models implies that neither should be favored in the absence of information other than obtained from the data. Nevertheless, both of them suggest that approximately 10% of the observed variety in the site popularity is due to mental states (and corresponding activities) most often experienced (pursued) by the consumers at the e-market. The latter does not appear implausible in the light of the Internet demographic survey for 1997 by Nielson Media Research (http://www.nielsenmedia.com) stating that 73% of the consumers used the WWW to search for information about products and services by means of accessing a small number of Web portals and search engines, such as Yahoo®, etc.

It is understood that for any "complete" validation of the proposed model, many more experiments are required but are beyond the limits of this paper. Additional cross-checking and verification are, however, still indispensable because technically, derivation of the Lomax (Pareto Second Kind or General Pareto) distribution as a gamma mix of exponentials was first reported several decades ago (Harris, 1968) but did not receive due attention in complex system research.

One supportive argument for the proposed approach is that it does not contradict the findings about the dynamics of social networks reported in the literature, but instead generalizes them. The widely held form of the power law $P(k) = \nu b^\nu / k^{\nu+1}$ can be obtained from Equation (8) for $k \gg b$ by Taylor series expansion (the minuend – by small $(b-1)/k$, and the subtrahend – by small $b/k$) under the assumption that the investigated system is homogeneous (i.e. by setting $M = 1$).

Let us now consider an asymptotic approximation of the model, which is also the continuous counterpart of Equation (8):

$$P_c(k) = \nu b^\nu (k+b)^{-\nu-1}. \qquad (9)$$

From

$$R(x) = l \int_x^\infty P_c(k) dk = l b^\nu (b+x)^{-\nu}, \qquad (10)$$

where $R(x)$ gives $r$ the rank of a unit of size $x$, $x_{(1)} \geq x_{(2)} \geq \cdots \geq x_{(r)} \geq \cdots \geq x_{(l)}$, in a set of $l$ objects, one can easily obtain

$$f_r = b l^{-1+1/\nu} r^{-1/\nu} - b l^{-1}, \qquad (11)$$

where $f_r$ is the relative occurrence frequency of the $r$-th popular unit. When $r \gg u$, this result coincides with the empirical formula $f_r = d + q(u+r)^{-\alpha}$, $d$, $q$, and $u$ are some constants, obtained by the authors of Reference (Krashakov *et al.*, 2006). Moreover, the negative values of $d$ empirically calculated in the latter study are generally in agreement with what would be estimated based on the sample size by applying formula (11), where $b, l > 0$ by definition.

As a final remark, let us mention that a lognormal distribution is often discussed as an alternative to the power law when describing the dynamics of complex systems (Mitzenmacher, 2003; Stouffer *et al.*, 2005). Given $k \gg b$ and some $m$, $2m \gg \ln k$, Equation (8) can be approximated as

$$P(k) = \frac{\nu b^\nu e^{\frac{\nu m}{2}}}{k} e^{-\frac{(\ln k + m)^2}{2m/\nu}} \qquad (12)$$

that gives its lognormal asymptotic form.

## 6 CONCLUSIONS

Having defined the overall goal as to deliver a universal but simple and accurate theoretical model for the observed behavior of a large class of complex systems, in this particular paper we focused on the formation of the dynamics of social networks and on methods for the network structure analysis. A mathematical model correctly describing these phenomena would help optimize recourse and service allocation as well as economic and management policies for companies in both the traditional and electronic business sectors, and also for organizations involved in collaborative activities, such as distribution of funds, innovation and know-how exchange, and so on.

We have applied the apparatus of statistical physics to describe the emergence of social networks. The network dynamics was defined in terms of its structure (i.e. how many subsystems are there and what is, as observed, their influence on the overall dynamics) as well as parameters of its elementary constituents (these parameters are the mental reaction time and, possibly, response times of external systems coupled with or simply affecting the social network). In the presented experiments, the proposed model has demonstrated a prognostic potential far superior to any of the classical modeling approaches. At the same time, the model proved to be quite encompassing but natural and thus easy to interpret and validate.

In our prior research reported elsewhere, the system-theoretic framework was successfully applied to capture the structure of different languages and to compare the efficiencies of text- and hypermedia- based communication (Kuleshov *et al*., 2005; Kryssanov *et al*., 2005). In future studies, we plan to apply the developed approach to explore article and patent authorship networks.

## ACKNOWLEGEMENT

The authors would like to thank, without implicating, Lada A. Adamic and Jean-Pierre Eckmann for providing the data used in the research.